\documentclass{PoS}

\title{Fermionic fields in the pseudoparticle approach}

\ShortTitle{Fermionic fields in the pseudoparticle approach}

\author{\speaker{Marc Wagner}\\
        Humboldt-Universit\"at zu Berlin, Institut für Physik, Newtonstra{\ss}e 15, D-12489 Berlin, Germany\\
        E-mail: \email{mcwagner@physik.hu-berlin.de}}

\abstract{
The pseudoparticle approach is a numerical method to compute path integrals without discretizing spacetime. The basic idea is to consider only those field configurations, which can be represented as a linear superposition of a small number of localized building blocks (pseudoparticles), and to replace the functional integration by an integration over the pseudoparticle degrees of freedom. In previous papers we have successfully applied the pseudoparticle approach to SU(2) Yang-Mills theory. In this work we discuss the inclusion of fermionic fields in the pseudoparticle approach. To test our method, we compute the phase diagram of the 1+1-dimensional Gross-Neveu model in the large-$N$ limit as well as the chiral condensate in the crystal phase.
}

\FullConference{The XXV International Symposium on Lattice Field Theory\\
		 July 30 - August 4 2007\\
		 Regensburg, Germany}

\newcommand{\ltapprox}{\raisebox{-0.5ex}{$\,\stackrel{<}{\scriptstyle\sim}\,$}}

\begin{document}


\section{Introduction}

Recently there have been several papers proposing models for SU(2) Yang-Mills theory with a small number of physically relevant degrees of freedom. These models include ensembles of regular gauge instantons and merons \cite{Lenz:2003jp,Lenz:2007st}, the pseudoparticle approach \cite{Wagner:2005vs,Wagner:2006qn,Wagner:2006du}, superpositions of calorons with non-trivial holonomy \cite{Gerhold:2006sk,Gerhold:2006kw} and an ensemble of dyons \cite{Diakonov:2007nv}. The common basic principle is to restrict the Yang-Mills path integral to those gauge field configurations, which can be represented as a linear superposition of a small number of localized building blocks (pseudoparticles), e.g.\ instantons, merons, akyrons, calorons or dyons.

These models have been quite successful, when dealing with problems related to confinement. First of all, the potential of two static charges is essentially linear within phenomenologically relevant distances. Moreover, a confinement-de\-confinement phase transition can be modeled, and numerical results for various quantities, e.g.\ the string tension, the topological susceptibility, the critical temperature or the low lying glueball spectrum, are in qualitative agreement with results from lattice calculations.

However, all these models exclusively consider pure Yang-Mills theory. Therefore, incorporating fermions is an interesting issue. In this paper we present first steps in this direction: we propose a method how to deal with fermionic fields in the pseudoparticle approach, and we test this method by applying it to a simple interacting fermionic theory, the $1$+$1$-dimensional Gross-Neveu model in the large-$N$-limit.


\section{\label{SEC001}Fermionic fields in the pseudoparticle approach}


\subsection{Basic principle}

The starting point is action and partition function of any theory with quadratic fermion interaction:
\begin{eqnarray}
 & & \hspace{-0.7cm} S[\psi,\bar{\psi},\phi] \ \ = \ \ \int d^{d+1}x \, \Big(\bar{\psi} Q(\phi) \psi + \mathcal{L}(\phi)\Big) \\
 & & \hspace{-0.7cm} Z \ \ = \ \ \int D\psi \, D\bar{\psi} \, \int D\phi \, e^{-S[\psi,\bar{\psi},\phi]} ,
\end{eqnarray}
where $\phi$ denotes any type and number of bosonic fields, e.g.\ the non-Abelian gauge field in QCD, and $Q$ is the Dirac operator, which, of course, depends on these bosonic fields.

To stay close to the spirit of the pseudoparticle approach, we consider fermionic field configurations $\psi$, which can be represented as a linear superposition of a fixed number of pseudoparticles:
\begin{eqnarray}
\psi(x) \ \ = \ \ \sum_j \underbrace{\eta_j G_j(x)}_{\textrm{\tiny $j$-th pseudoparticle}} .
\end{eqnarray}
Each pseudoparticle is a product of a Grassmann valued spinor $\eta_j$ and a function $G_j$, which is localized in space as well as in time (the term pseudoparticle refers to this localization). The integration over all fermionic field configurations is defined as the integration over the Grassmann valued spinors $\eta_j$:
\begin{eqnarray}
\int D\psi \, D\bar{\psi} \, \ldots \ \ = \ \ \int \bigg(\prod_j d\eta_j \, d\bar{\eta}_j\bigg) \ldots
\end{eqnarray}
Integrating out the fermions yields
\begin{eqnarray}
\label{EQN001} & & \hspace{-0.7cm} S_\textrm{\tiny effective}[\phi] \ \ = \ \ \int d^{d+1}x \, \mathcal{L}(\phi) - \ln\Big(\det\Big(\langle G_j | Q | G_{j'} \rangle\Big)\Big) \\
 & & \hspace{-0.7cm} Z \ \ \propto \ \ \int D\phi \, e^{-S_\textrm{\tiny effective}[\phi]} ,
\end{eqnarray}
where the ``fermionic matrix'' $\langle G_j | Q | G_{j'} \rangle$ is the Dirac operator represented in the pseudoparticle basis. We will refer to this pseudoparticle regularization as $Q$-regularization, and we will shortly point out that this $Q$-regularization is not suited to produce physically meaningful results.

In the case that $\det(Q)$ is real and positive, $\det(Q) = \sqrt{\det(Q^\dagger Q)}$. This suggests another pseudoparticle regularization:
\begin{eqnarray}
\label{EQN002} S_\textrm{\tiny effective}[\phi] \ \ = \ \ \int d^{d+1}x \, \mathcal{L}(\phi) - \frac{1}{2} \ln\Big(\det\Big(\langle G_j | Q^\dagger Q | G_{j'} \rangle\Big)\Big) .
\end{eqnarray}
In the following section we will argue that this $Q^\dagger Q$-regularization has significant advantages over the $Q$-regularization (\ref{EQN001}).

Note that using eigenfunctions of the Dirac operator as ``pseudoparticles'' yields the well known finite mode regularization \cite{Andrianov:1982sn,Andrianov:1983fg}.

\subsection{The $Q$-regularization versus the $Q^\dagger Q$-regularization}

The problem of the $Q$-regularization (\ref{EQN001}) is that applying the Dirac operator $Q$ to one of the pseudoparticles $G_{j'}$ in general yields a function, which is partially outside the pseudoparticle function space $\textrm{span}\{G_n\}$:
\begin{eqnarray}
Q G_{j'}(x) \ \ = \ \ \sum_k a_{j' k} G_k(x) + h_{j'} H_{j'}(x)
\end{eqnarray}
with $H_{j'}$ normalized and $H_{j'} \perp \textrm{span}\{G_n\}$. If $|\sum_k a_{j' k} G_k| \gg |h_{j'}|$, the situation is uncritical. However, as soon as $|\sum_k a_{j' k} G_k| \ltapprox |h_{j'}|$, serious problems arise: when computing the fermionic matrix elements $\langle G_j | Q | G_{j'} \rangle$, a significant part of $Q G_{j'}$ is simply ignored, namely $h_{j'} H_{j'}$, because it is perpendicular to the pseudoparticle function space $\textrm{span}\{G_n\}$.

On the other hand, the $Q^\dagger Q$-regularization (\ref{EQN002}) has the following advantage: both the left hand sides $\langle G_j | Q^\dagger$ and the right hand sides $Q | G_{j'} \rangle$ of the fermionic matrix elements $\langle G_j | Q^\dagger Q | G_{j'} \rangle$ might be (partially) outside to the pseudoparticle function space $\textrm{span}\{G_n\}$, but they form the same function space $\textrm{span}\{Q G_n\}$, in which their overlap is computed. Of course, the above problem of partially perpendicular left and right hand side function spaces does not exist anymore.

For more elaborate arguments, especially why one can expect to obtain correct results from the $Q^\dagger Q$-regularization, we refer to \cite{Wagner:2007he}.


\section{Testing the method: the Gross-Neveu model in the pseudoparticle approach}


\subsection{The $1$+$1$-dimensional Gross-Neveu model in the large $N$-limit}

As a testbed for our pseudoparticle method we use the Gross-Neveu model \cite{Gross:1974jv}, which is a four fermion interacting theory with $N$ identical flavors. Action and partition function of the $1$+$1$-dimensional Gross-Neveu model are given by
\begin{eqnarray}
 & & \hspace{-0.7cm} S \ \ = \ \ \int d^2x \, \bigg(\sum_{n=1}^N \bar{\psi}^{(n)} \Big(\gamma_0 (\partial_0 + \mu) + \gamma_1 \partial_1\Big) \psi^{(n)} - \frac{g^2}{2} \bigg(\sum_{n=1}^N \bar{\psi}^{(n)} \psi^{(n)}\bigg)^2\bigg) \\
 & & \hspace{-0.7cm} Z \ \ = \ \ \int \bigg(\prod_{n=1}^N D\psi^{(n)} \, D\bar{\psi}^{(n)}\bigg) e^{-S} ,
\end{eqnarray}
where $\mu$ is the chemical potential and $g$ the dimensionless coupling constant. To get rid of the four fermion term, one usually introduces a scalar field $\sigma$. Integrating out the fermions yields
\begin{eqnarray}
\label{EQN003} & & \hspace{-0.7cm} S_\textrm{\tiny effective} \ \ = \ \ N \left(\frac{1}{2 \lambda} \int d^2x \, \sigma^2 - \ln\Big(\det\Big(\gamma_0 (\partial_0 + \mu) + \gamma_1 \partial_1 + \sigma\Big)\Big)\right) \\
\label{EQN004} & & \hspace{-0.7cm} Z \ \ \propto \ \ \int D\sigma \, e^{-S_\textrm{\tiny effective}}
\end{eqnarray}
with $\lambda = N g^2$.

In the following we consider the large-$N$ limit, in which the model can be solved analytically \cite{Dashen:1974xz,Wolff:1985av,Schnetz:2004vr}. This amounts to using an infinite number of flavors $N$, while $\lambda = N g^2$ is kept constant. Note that in the $N \rightarrow \infty$ limit only a single $\sigma$-field configuration contributes to the partition function (\ref{EQN004}) minimizing the effective action. Note also that in the large-$N$ limit $\sigma$ is proportional to the chiral condensate, i.e.\ $\sigma = -g^2 \sum_{n=1}^N \bar{\psi}^{(n)} \psi^{(n)}$.


\subsection{Numerical results: the phase diagram and the chiral condensate}

From a technical point of view computations in the pseudoparticle approach are quite similar to those in lattice field theory. The number of pseudoparticles corresponds to the number of lattice sites, while the distance between neighboring pseudoparticles plays a role similar to the lattice spacing. The scale can be set by any dimensionful quantity and it can be changed by choosing a different value for the dimensionless coupling constant. For a recent lattice study of the Gross-Neveu model we refer to \cite{de Forcrand:2006ut}.

\begin{figure}[b!]
\begin{center}
\includegraphics{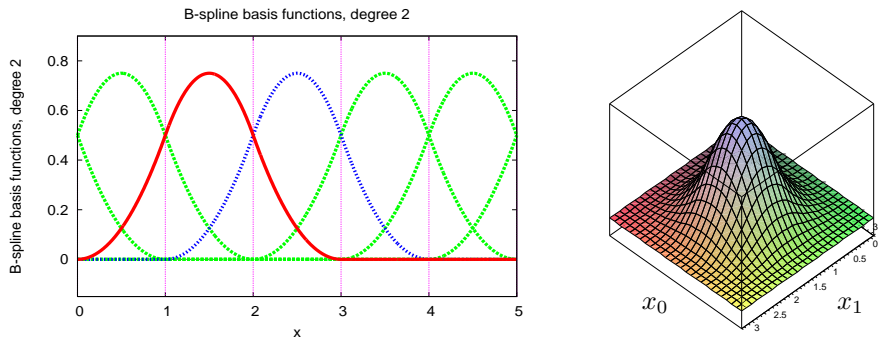}
\caption{\label{FIG001}B-spline basis functions in one and two dimensions.}
\end{center}
\end{figure}

For the following computations we apply the $Q^\dagger Q$-regularization (\ref{EQN002}). As pseudoparticles we use a large number of uniformly distributed hat functions, more precisely B-spline basis functions of degree $2$ (cf.\ e.g.\ \cite{Wolf06}), which are shown in Figure~\ref{FIG001}. There is one fermionic pseudoparticle per unit volume, the spatial extension of the periodic spacetime region is $L_1 = 144$ and the temporal extension $L_0$ varies, corresponding to different temperatures $T = 1/L_0$. The main reason for considering such pseudoparticles is that they yield a sensible set of field configurations: they form a piecewise polynomial basis of degree $2$, i.e.\ any not too heavily oscillating field configuration can be approximated. Therefore, if the pseudoparticle method we have presented in Section~\ref{SEC001} is a useful numerical technique, we can expect to reproduce correct Gross-Neveu results. In other words, B-spline basis functions are suitable pseudoparticles for testing our approach.

\begin{figure}[b!]
\begin{center}
\includegraphics{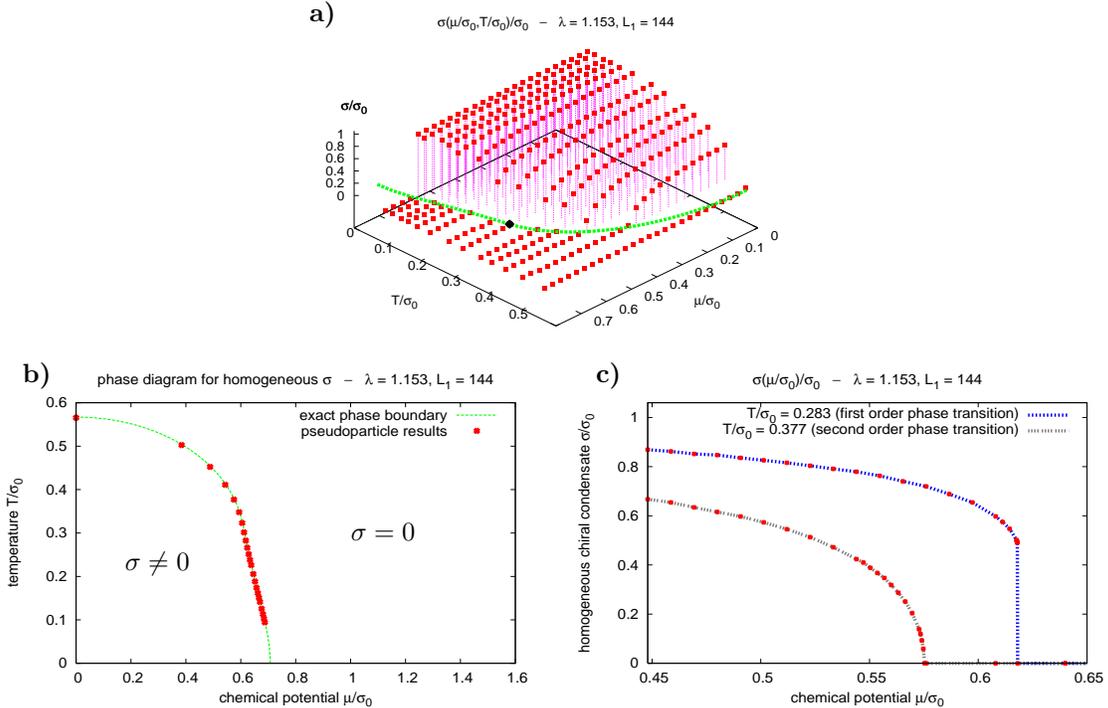}
\caption{\label{FIG002}
\textbf{a)}~$\sigma/\sigma_0$ as a function of $\mu/\sigma_0$ and $T/\sigma_0$ (red dots) together with the analytically obtained phase boundary (blue line) and the tricritical point $(\mu/\sigma_0,T/\sigma_0) = (0.608,0.318)$ separating first and second order phase transitions (black dot).
\textbf{b)}~Phase diagram for homogeneous chiral condensate (red dots: pseudoparticle results; green line: analytical result).
\textbf{c)}~Two sections trough the phase diagram showing $\sigma/\sigma_0$ as a function of $\mu/\sigma_0$ at $T/\sigma_0 = 0.283$ (first order phase transition) and $T/\sigma_0 = 0.377$ (second order phase transition).
}
\end{center}
\end{figure}

At first we perform computations of the chiral condensate $\sigma$ at chemical potential $\mu = 0$ and temporal extension $L_0 = 8$ for various values of the coupling constant $\lambda$. As it is in lattice calculations different values of $\lambda$ correspond to different physical extensions of the spacetime region and, therefore, to different values of the temperature. From these computations we determine that value of $\lambda$, where $\sigma$ just vanishes: $\lambda_\textrm{\tiny critical} = 1.153$. For all further computations we use $\lambda = \lambda_\textrm{\tiny critical}$. By doing this we have set the scale, since from now on $L_0$ plays the role of inverse temperature such that $L_0 = 8$ corresponds to the critical temperature of chiral symmetry breaking.

After that, we perform a low temperature computation at $L_0 = 48$ or equivalently $T = T_\textrm{\tiny critical} / 6$, to obtain an approximation of the zero temperature value of the chiral condensate: $\sigma_0 = 0.221$. This allows us to express all dimensionful quantities in terms of $\sigma_0$.

\begin{figure}[b!]
\begin{center}
\includegraphics{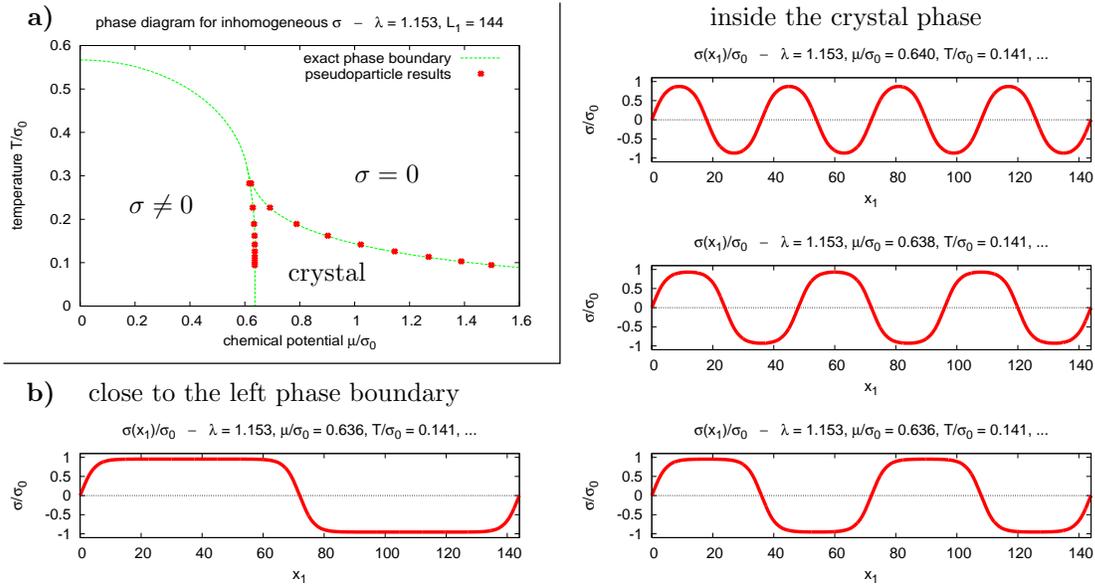}
\caption{\label{FIG003}
\textbf{a)}~Phase diagram for inhomogeneous chiral condensate (red dots: pseudoparticle results; green line: analytical result).
\textbf{b)}~The pseudoparticle chiral condensate for $T/\sigma_0 = 0.141$ and different values of $\mu/\sigma_0$.
}
\end{center}
\end{figure}

Now we are in a position to compute the chiral condensate at arbitrary temperature $T/\sigma_0$ and chemical potential $\mu/\sigma_0$. Results for homogeneous chiral condensate are shown in Figure~\ref{FIG002}a together with the analytically obtained phase boundary \cite{Dashen:1974xz,Wolff:1985av} and the tricritical point separating first and second order phase transitions. Pseudoparticle and analytical results are in excellent agreement both for the phase boundary (cf.\ also Figure~\ref{FIG002}b) and for the order of the phase transition (cf.\ also Figure~\ref{FIG002}c, where we have plotted $\sigma/\sigma_0$ as a function of $\mu/\sigma_0$ for two different values of $T/\sigma_0$, one in the first order region and the other in the second order region).

For inhomogeneous chiral condensate a third so called crystal phase appears \cite{Schnetz:2004vr}, where the minimum of the effective action (\ref{EQN003}) is not anymore given by a homogeneous chiral condensate $\sigma$. In addition to the fermionic fields we also represent $\sigma$ in terms of B-spline pseudoparticles (for details cf.\ \cite{Wagner:2007he}). As before, the pseudoparticle phase diagram and the analytically obtained phase diagram are essentially indistinguishable (cf.\ Figure~\ref{FIG003}a).

We have also compared the pseudoparticle chiral condensate and the analytically obtained chiral condensate at various points $(\mu/\sigma_0 , T/\sigma_0)$ inside the crystal phase; again, there is excellent agreement. Figure~\ref{FIG003}b shows the emergence of a crystalline structure: the kink-antikink structure close to the left phase boundary changes to a sin-like behavior, when approaching the center of the crystal phase.

Note that we have performed the same computations also with the naive $Q$-regularization. As expected the results are completely wrong, e.g.\ there is no chirally symmetric phase even in the simple case of homogeneous chiral condensate. One can easily show that this is inherent to the $Q$-regularization and not a problem of the number or the type of pseudoparticles applied \cite{Wagner:2007he}.


\section{Summary and outlook}

We have proposed a method to incorporate fermionic fields in the pseudoparticle approach. While the naive $Q$-regularization is not suited to produce any useful results, the $Q^\dagger Q$-regularization has the potential to yield correct and physically meaningful results. The computation of the phase diagram of the Gross-Neveu model with the $Q^\dagger Q$-regularization both for homogeneous and for inhomogeneous chiral condensate has been a first successful test of the pseudoparticle approach applied to fermionic theories.

The next step is to apply the pseudoparticle approach to QCD and to identify a small number of physically relevant degrees of freedom, probably fermionic pseudoparticles, which are able to approximate typical low lying eigenmodes of the Dirac operator. The goal is to obtain a model with a small number of degrees of freedom, which exhibits both chiral symmetry breaking and a confinement deconfinement phase transition at the same time.


\begin{acknowledgments}

I would like to thank Martin Ammon, Ernst-Michael Ilgenfritz, Felix Karbstein, Frieder Lenz, Michael M{\"u}ller-Preussker, Jan Martin Pawlowski, Michael Thies and Konrad Urlichs for helpful discussions and useful comments.

\end{acknowledgments}



\end{document}